# *Operando analysis of a solid oxide fuel cell by environmental transmission electron microscopy*

Q. Jeangros,[1] M. Bugnet,[2] T. Epicier,[2,3] C. Frantz,[4] S. Diethelm,[4] D. Montinaro,[5] E. Tyukalova,[6] Y. Pivak,[7] J. Van herle,[4] A. Hessler-Wyser,[1] M. Duchamp[6,8]

[1] Photovoltaics and Thin-Film Electronics Laboratory (PVLab), École Polytechnique Fédérale de Lausanne (EPFL), Rue de la Maladière 71b, 2002 Neuchâtel, Switzerland.

[2] Univ Lyon, CNRS, INSA-Lyon, UCBL, MATEIS, UMR 5510, 69621 Villeurbanne, France

[3] Univ Lyon, UCBL, IRCELYON, UMR CNRS 5256, 69626 Villeurbanne, France

[4] Group of Energy Materials (GEM), École Polytechnique Fédérale de Lausanne (EPFL), Rue de l'Industrie 17, 1951 Sion, Switzerland.

[5] SOLIDpower S.p.A., 38017, Mezzolombardo, Italy

[6] Laboratory for *in situ* & *operando* Electron Nanoscopy, School of Materials Science and Engineering, Nanyang Technological University, 50 Nanyang Avenue, Singapore 639798.

[7] DENSsolutions, Informaticalaan 12, 2628 ZD Delft, The Netherlands

[8] MajuLab, International Joint Research Unit UMI 3654, CNRS, Université Côte d'Azur, Sorbonne Université, National University of Singapore, Nanyang Technological University, Singapore, Singapore

**Abstract**

Correlating the microstructure of an energy conversion device to its performance is often a complex exercise, notably in solid oxide fuel cell (SOFC) research. SOFCs combine multiple materials and interfaces that evolve in time due to high operating temperatures and reactive atmospheres. We demonstrate here that *operando* environmental transmission electron microscopy can simplify the identification of structure-property links in such systems. By contacting a cathode-electrolyte-anode cell to a heating and biasing microelectromechanical system in a single-chamber configuration, a direct correlation is found between the environmental conditions ($O_2$ and $H_2$ partial pressures, temperature), the cell voltage, and the microstructural evolution of the fuel cell, down to the atomic scale. The results shed new insights into the impact of the anode oxidation state and its morphology on the cell electrical properties.

**Introduction**

Improving the performance and stability of renewable energy conversion technologies generally requires inputs provided by advanced microstructural characterization methods. Techniques based on electrons or X-rays can provide detailed insights into atomic organization, chemistry, and overall microstructure down to the atomic scale. However, these analyses are typically performed *ex situ* or at best *in situ*, hence complicating the identification of structure-property links. Indeed, analysis conditions often differ from the ones experienced by the materials in a functioning device, which may lead to microstructural alterations, and the macroscale metric of interest, *e.g.*, the electrical properties of the device, is not recorded during the analysis at pertinent spatial resolution.

The analysis of the microstructure of materials during operation is particularly challenging for solid oxide fuel cells (SOFCs) and their solid oxide electrolysis cells (SOECs) counterparts. Their harsh operating conditions combining high temperatures (600-1000 °C to ensure sufficient ionic conductivity of the electrolyte),[1] reducing and oxidising gases (typically $H_2$ and $O_2$ or air), and electrical bias are difficult to recreate within characterisation setups. Overall, SOFCs convert the chemical energy of a fuel directly into electricity through an electrochemical process or, vice versa, electricity into usable and storable fuels via SOECs.[2,3] This technology is likely to play a key role in the transition towards renewable energy sources due to its versatility, negligible emissions of $NO_x$ or $SO_x$, and high performance. Standard SOFC/SOEC designs include an yttria-stabilized zirconia (YSZ) electrolyte sandwiched between a thick nickel/YSZ anode support and a cathode based on strontium-doped lanthanum manganite (LSM) or lanthanum strontium cobalt ferrite (LSCF), the latter requiring the use of a ceria diffusion barrier.[2,4] For the anode, NiO is typically co-sintered with YSZ and then reduced to its metallic Ni active state during the first operation of the cell, *i.e.,* when the temperature is increased up to the operating setpoint and the fuel is introduced on the anode side. The volume loss associated with this reduction reaction leaves pores in the anode, ensuring a permeation of the fuel to the electrochemically active sites, the triple-phase boundaries (TPBs, Ni-YSZ-porosity on the anode side).[5] Some degradation mechanisms may then occur during long-term operation as a result of the high operating temperatures.[6] TPBs in both anode and cathode (LSM-YSZ-porosity) may deactivate as a result of various poisoning mechanisms or microstructural alterations.[5,7–16]

Due to its ability to retrieve chemical and crystallographic properties at high spatial resolution, electron microscopy has been one method of choice to identify degradation mechanisms in SOFCs,[10,17–21] notably when applied *in situ* by raising temperature in a relevant gas atmosphere directly inside an environmental electron microscope.[22–29] For example, environmental transmission electron microscopy (ETEM) coupled to various spectroscopies has enabled a direct visualisation of the reduction and reoxidation pathways of the Ni anode catalyst.[27–32] Indeed, modern aberration-corrected environmental transmission electron microscopes now achieve a spatial resolution below 0.1 nm and enable a direct visualisation of solid-gas interactions down to the atomic scale. A few recent examples include the direct observation of metal sintering,[33] metal nanoparticles-support interactions,[34] surface reconstructions,[35] atomic-scale dynamics,[36,37] or phase transformations.[38] However, previous ETEM studies on SOFCs and other types of systems have generally focused on one specific aspect of the technology of interest. For example the evolution of the SOFC anode was investigated at high temperature and in a reactive gas atmosphere,[27–31] yet without providing a complete picture of the operation of the device since its electric properties were not measured at the same time in the microscope.

Overall, the absence of a comprehensive *operando* TEM investigation of SOFCs (and of other energy conversion technologies) stems from the complexity of combining and controlling all the required stimuli at the same time on the same sample in the microscope (here high temperatures, reducing/oxidizing atmosphere and electrical bias). Capitalizing on recent advances in focused ion beam (FIB) sample preparation protocols,[39–42] in microelectromechanical systems (MEMS) for combined heating and biasing studies inside microscopes,[43,44] and in ETEM techniques,[45] we demonstrate that SOFCs can be analysed *operando*. To do so, we measure the voltage of a cell exposed simultaneously to $H_2$ and $O_2$ at a sufficiently high temperature (600 °C) and at the same time monitor its microstructure down to the atomic scale. A single-chamber configuration, *i.e.*, where the entire SOFC sample and both oxidant/reducing gases are present in the same chamber, is selected to avoid the need to constrain the oxidant gas to the cathode and the fuel to the anode. In these conditions, the difference in catalytic activity of the two electrodes leads to a voltage difference between the anode and cathode.[26,46] By varying the $O_2$-to-$H_2$ ratio, direct correlations between the composition of the gas mixture, the cell voltage and the anode microstructure are established, providing new insights into the operation mechanisms of SOFCs. These experiments open new perspectives for the analysis of the links between performance and microstructure of energy materials.

**Results and discussion**

The cell architecture investigated here consists of an LSM/YSZ cathode, a YSZ electrolyte and a NiO/YSZ anode, as shown in Figure 1. To ensure industrial relevance, the SOFC investigated here was produced by SOLIDpower S.p.A. using tape-casting. The electrolyte was made thinner (2 µm) to enable the fabrication by FIB of a TEM lamella containing all the relevant interfaces of the cathode-electrolyte-anode cell. The lamella was contacted to a MEMS chip from DENSsolutions with simultaneous heating and biasing capabilities (Figure 1a, see Materials & Methods for details). The TEM lamella was mounted onto a prototype DENSsolutions MEMS holder and inserted in the column of a FEI Titan G2 ETEM. A scanning TEM (STEM) annular dark-field image (ADF) image of the as-prepared SOFC is shown in Figure 1b. Corresponding elemental maps obtained by STEM energy dispersive X-ray spectroscopy (EDX) are displayed in Figure 1c, highlighting how the different phases are distributed in the initial sample. The YSZ electrolyte is dense with grains of about 1-2 µm, while the LSM/YSZ cathode is porous to ensure air access to the TPBs (LSM-YSZ-porosity on the cathode side). On the other hand, the NiO/YSZ anode precursor side is denser in its as-sintered state.

Prior to operation, the as-sintered NiO phase needs to be reduced to Ni, its active state. The porosity that will result from the process will also enable to form TPBs on the anode side.[5] To activate the Ni catalyst of the anode cermet, 10 to 15 mbar of forming gas (5 v/v% of $H_2$ in $N_2$) was introduced in the ETEM, a pressure close to the maximal pressure allowed in the environmental chamber. The temperature was increased up to 750 ºC to trigger the reduction of NiO to Ni. STEM ADF micrographs acquired at various temperatures and pressures are given in Figure S1, highlighting how the microstructural changes occur in the anode during the activation of the Ni catalyst. The reduction reaction becomes visible through the creation of pores within the NiO grains, with pores forming preferentially at the interfaces with YSZ due to a quick coarsening of the Ni phase at these temperatures. As highlighted in prior studies,[27,28,30,32] the NiO reduction kinetics is slow inside the ETEM in these flow and pressure conditions. The reaction rate is initially controlled by the nucleation of the first Ni seeds. The presence around the reaction sites of $H_2O$ released by the reduction then likely slows down the reaction rate at high conversion fractions. On the other hand, the cathode remains unchanged in forming gas/$H_2$ up to this temperature of 750 °C and within this time scale of 210 minutes (Figure S2).

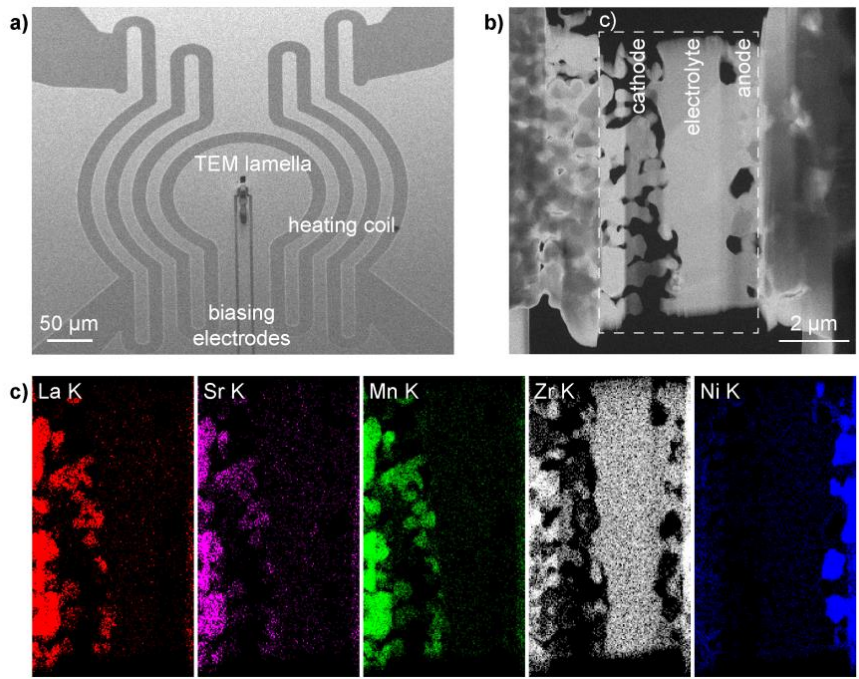

*Figure 1: a) SEM image of a biasing and annealing MEMS chip for operando TEM. The anode and cathode of the SOFC lamella are electrically connected to the biasing electrodes of the MEMS. b) STEM ADF micrograph of the electrically connected SOFC sample, and c) corresponding STEM EDX maps of the initial SOFC device acquired from the dashed area shown in b).*

To trigger the oxidation of $H_2$ at the anode and the reduction of $O_2$ at the cathode and hence the operation of the SOFC lamella in a single-chamber configuration, the temperature was lowered to 600 °C (notably to slow down the Ni oxidation process). The forming gas flow was then set to 3 ml/min before introducing an additional flow of $O_2$ of ~0.1 ml/min, leading to an increasing $O_2$-to-$H_2$ ratio in the ETEM. Note that the $O_2$ flow was set to the minimum value allowed by the Brooks mass flow controller ahead of the ETEM. In these conditions, the total pressure in the ETEM chamber reached 15.8 mbar. $N_2$ was employed as a mixing agent for safety concerns due to the necessity of mixing $H_2$ and $O_2$ to trigger the operation of the SOFC in a single-chamber configuration.

We investigated the impact of a varying $O_2$-to-$H_2$ ratio and monitored the cell voltage in relation to the morphology of the Ni catalyst (Figures 2). Note that several reduction-oxidation cycles took place between Figures S1 and Figure 2. Figure 2a plots i) the evolution with time of the average TEM image intensity of the two Ni grains shown in Figures 2b-g (*i.e.*, intensity of the image averaged over the area of the two grains), ii) the ratio between the $O_2$ and $H_2$ signals obtained from the residual gas analyser (RGA) appended to the exit of the ETEM chamber, and iii) the voltage between the two MEMS biasing electrodes. The as-measured RGA $O_2$-to-$H_2$ ratio data (full line) was advanced by 180 seconds to correct for the time needed by the gas to travel from the reaction chamber to the RGA (dashed line, see supplementary information for details). Figure 2b-g shows a selection of TEM images detailing the evolution of the two Ni grains, the intensities of which are plotted in Figure 2a. A movie of the full sequence of TEM images of Figure 2 is shown in the supplementary information (Video S1). From Figure 2a, a direct correlation between Ni grain average intensity, presence of $O_2$, and voltage between the anode and cathode is noticed. When introducing $O_2$ in the ETEM chamber, the image intensity remains constant for about 500 s, which coincides with a small increase in voltage between the MEMS electrodes. As the $O_2$-to-$H_2$ ratio increases further (from 600 s to 1500 s), the cell voltage drops rapidly back to a value close to its initial baseline, while the Ni grains become darker. This lowering of TEM image intensity is indicative of an oxidation of the Ni grains to NiO: oxygen is incorporated in the Ni grains, leading to a thickening of the grains and to the filling of voids (see arrowheads in Figure 2b-d), which in turn decreases the number of electrons collected by the TEM camera due to additional scattering to high angles. This oxidation of the Ni catalyst in the TEM images is confirmed by tracking the evolution of electron energy-loss spectra (EELS) of the Ni-$L_{2,3}$ edges (Figure S3). The increasing intensity of the Ni-$L_3$ edge with respect to the $L_2$ edge indicates an oxidation of Ni during the first part of the experiment.[47] As discussed elsewhere,[31,48,49] this volume expansion of Ni upon oxidation is larger than that predicted by the Pilling-Bedworth ratio due to unbalanced mass transport mechanisms. In this temperature range <1000 °C, $Ni^{2+}$ ions diffuse outwards through the NiO scale grain boundaries faster than $O^{2-}$ ions diffuse inwards, leading to the injection of voids at the Ni/NiO interface and eventually to the formation of internal voids within the growing NiO scale. When stopping the $O_2$ flow at ~1500 s, a delay of several minutes (until ~2300 s) is observed before the image intensity starts to increase again as the NiO grains reduce back to Ni (Figure 2f). The Ni grains shrink during the reduction reaction and porosity re-appears within these grains (see arrowheads in Figure 2e-g). In parallel, the voltage starts to increase when $O_2$ is removed after ~1500 s, before decreasing from ~2500 s onwards. The voltage increases and decreases at a slower rate compared to the first peak (when $O_2$ was introduced in the chamber).

*Ex situ* tests made in an oven with the same materials, this time in the form of a 14-mm diameter button cell, reveal a similar trend: a peak in open circuit voltage at intermediate $O_2$-to-$H_2$ ratios and then a drop in voltage at higher ratios (Figure S4). A decrease in $O_2$-to-$H_2$ ratio then leads to another voltage peak at intermediate ratios, and finally to a sharp decrease in voltage as $O_2$ is fully removed from the chamber.

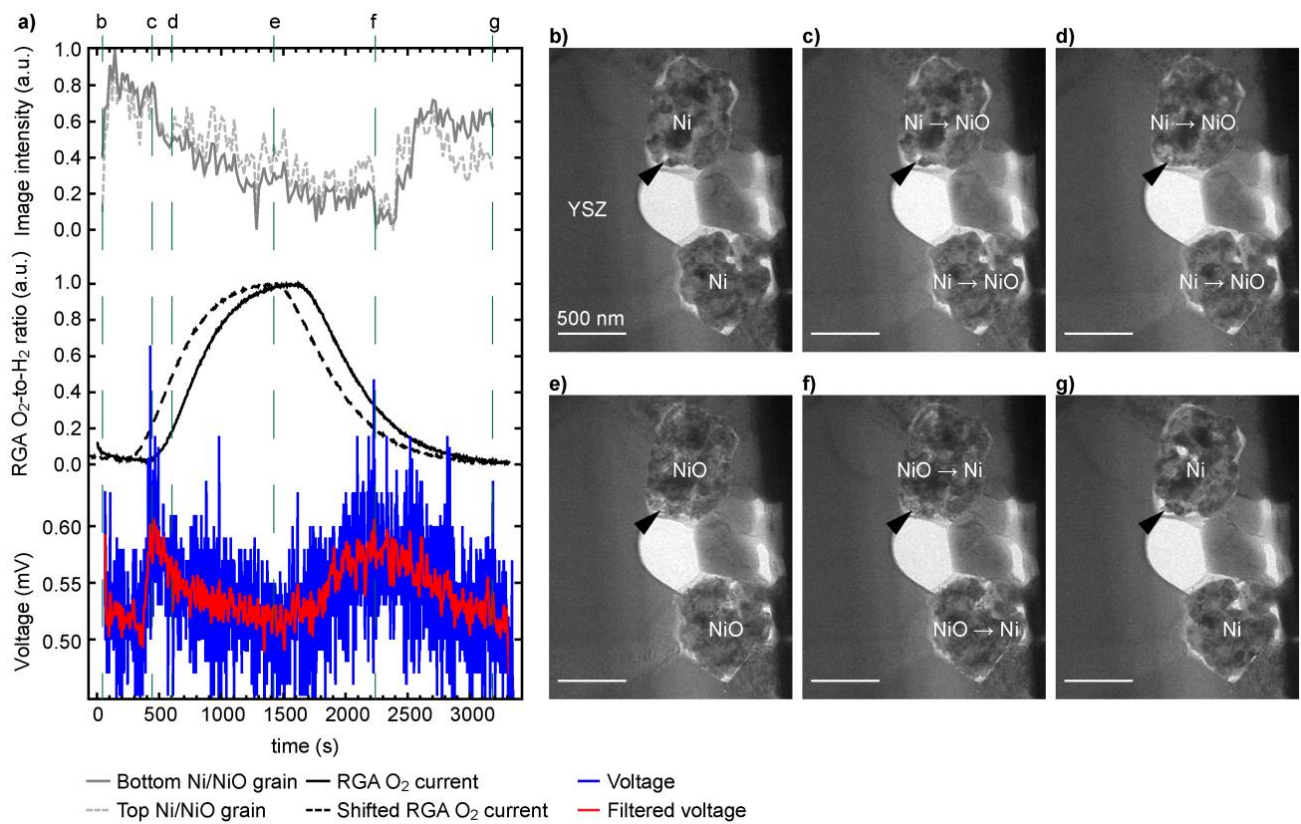

*Figure 2: a) Plots showing the evolution with time of the average TEM image intensity measured at the location of two Ni grains, the ratio of the RGA $O_2$ and $H_2$ signals (raw data, full line, and curve shifted forward by 180 seconds, dashed line), and the voltage measured between the two biasing electrodes (raw data in blue and after the application of a gaussian filter in red, see supplementary information for details). b-f) Selection of corresponding TEM images of the two Ni grains, the intensity of which is reported in a), taken at critical steps of the oxidation and reduction process. Arrowheads highlight morphological changes occurring during the oxidation and then reduction of Ni, see text for details.*

From Figure 2c and the EELS data of Figure S3, it appears that the first increase in voltage is correlated with the presence of Ni in its metallic state (smaller volume, compact morphology with some open porosity as shown by the arrowhead, lower Ni $L_3/L_2$ EELS ratio). This observation is consistent with the need to have Ni at the anode TPBs for the cell to operate and deliver a voltage. In these conditions, $H_2$ can adsorb and dissociate on Ni and then react at the TPBs with $O^{2-}$ coming from the cathode through the YSZ phase, leading to an open circuit voltage between the cathode and the anode. The voltage increase that we measure hence suggests that the cell is electrochemically active and is operating in a single-chamber configuration at certain $O_2$-to-$H_2$ ratios. The small voltage gain we measure (in the sub mV range) is orders of magnitudes below the open circuit voltage measured for the same combination of materials in a single-chamber configuration in an oven (0.8 V, see Figure S4). However, it should be mentioned that only few TPBs are present in our TEM lamella, and, as a result, the difference in oxygen partial pressure between the anode and the cathode is minute.[46] Based on the Nernst equation, voltage values of 0.1 mV correspond to a difference in $O_2$ partial pressure between anode and cathode of 0.5%. From the *ex situ* tests of Figure S4 and as-expected, the cell operates around the stoichiometric $O_2$-to-$H_2$ ratio of 0.5. Then, once the partial pressure of $O_2$ passes a certain threshold, Ni oxidizes to NiO, which stops the electrochemical reaction process, and the voltage drops (Figure 2d).

To rationalize the voltage variations observed in Figure 2 and explain the second voltage peak, similar sequences capturing the oxidation and reduction of the Ni catalyst were performed at higher spatial resolution. Figure 3 details the morphological changes occurring at the surface of one Ni grain during an oxidation and then a reduction. Figure 3a shows the evolution in time of an intensity profile taken across a Ni/void interface, which is shown in the form of a contour plot (taken from the region marked by an arrow in Figure 3b). The RGA and voltage data are also plotted in Figure 3a. A first increase in voltage is observed after ~380 s, which coincides with the presence of both $O_2$ and $H_2$ in the ETEM chamber and with Ni in its metallic state (as in Figure 2). Indeed, the intensity profile taken at the surface of one Ni grain does not change during these early stages, despite the (low) $O_2$ partial pressure now being present in the chamber (Figure 3a). The dense Ni grain morphology remains identical between Figure 3b and c. As the $O_2$-to-$H_2$ ratio increases after 400 s of experiment, a NiO scale starts to form on the metallic Ni grain (Figure 3d). The surface of the Ni grain retracts towards the centre of the Ni grain (see dashed line marking the Ni/NiO interface in Figure 3a, d-e). The Ni grain is now covered by a NiO scale that expands outwards. The TEM image intensity within the region that was previously a void now decreases as NiO is now forming there (arrowhead in Figure 3a and d). Once the $O_2$ flow is stopped and the $O_2$-to-$H_2$ ratio starts to decrease after 1500 s, the position of the Ni/NiO interface stops retracting and remains immobile along the y axis. In parallel, the intensity at the location of the NiO scale starts to decrease further (see black arrow in Figure 3a and e). As it will be confirmed below in Figure 4, this loss in intensity results from the growth of new Ni domains directly on the NiO scale as the $O_2$-to-$H_2$ ratio decreases. The voltage increases in these conditions, which would be consistent with the presence of Ni on the outer surface of the NiO scale to enable the oxidation of $H_2$. Furthermore, Ni $L_{3,2}$ EELS data shown in Figure S3 is consistent with such a mixed NiO/Ni system: an intermediate $L_3/L_2$ ratio is measured in these conditions. With such a morphology, $H_2$ can adsorb and dissociate on Ni, and if in contact with YSZ, contribute to the overall electrochemical reaction. After 2200 s, the Ni/NiO interface is observed to move downwards as the NiO scale disappears and the Ni islands present on the scale surface merge with the Ni grain (see half black half white arrows in Figure 3a and f). In Figure 3g, the NiO scale has completely disappeared. The second voltage increase from ~1500 s to ~2300 s is broader than the first one. The Ni sites at the anode TPBs remain electrochemically active as long as a sufficient $O_2$ partial pressure is present to sustain the $O_2$ reduction reaction on the cathode side. A movie of the full sequence of data shown in Figure 3 is shown in the supplementary information Video S2.

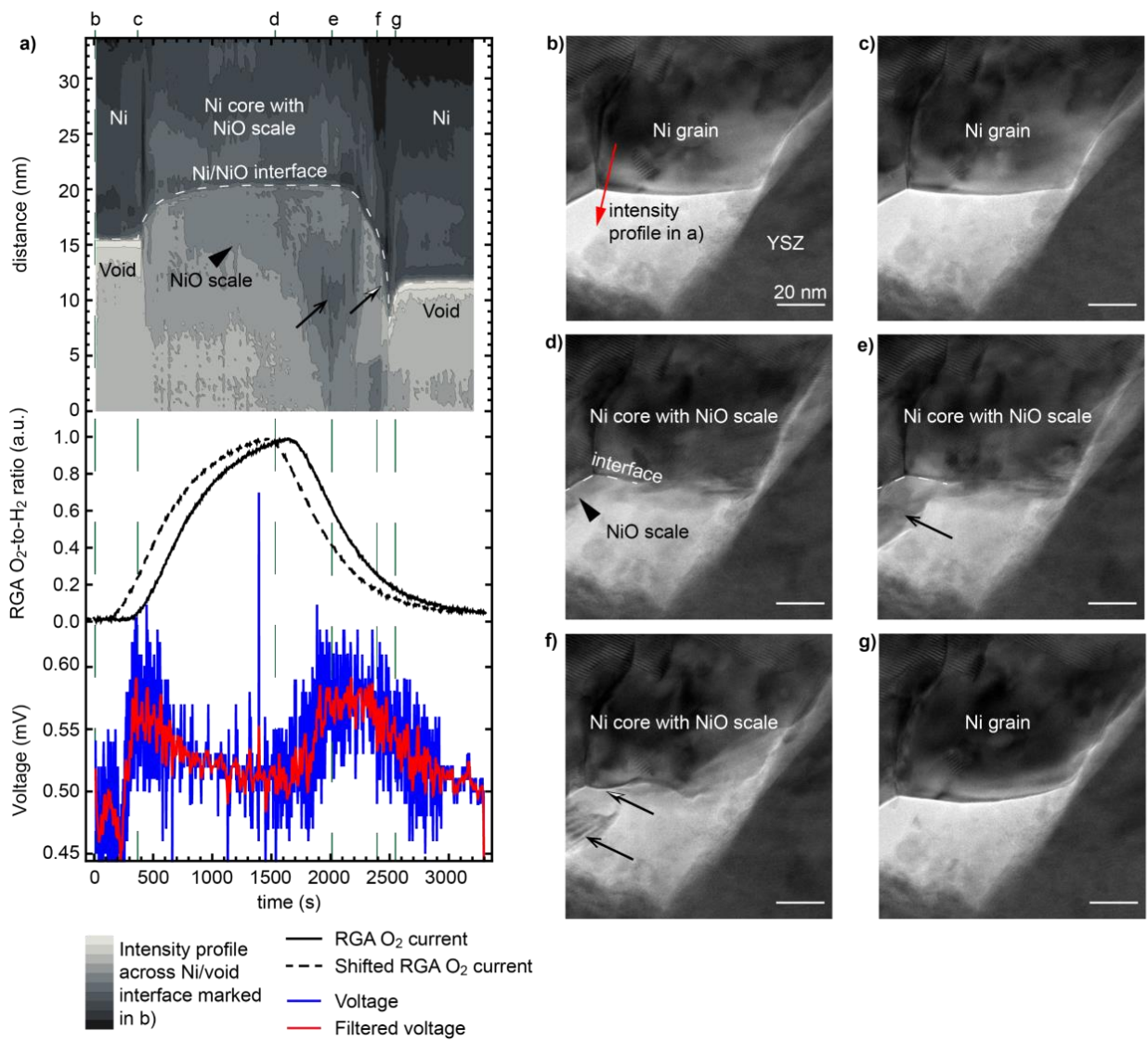

*Figure 3: Contour plot of the evolution of the TEM image intensity taken along the arrow shown in b), the RGA $O_2$-to-$H_2$ ratio (raw data, full line, and shifted forward by 180 seconds, dashed line), and voltage measured between the anode and cathode (raw and gaussian-filtered data). b-f) Selection of TEM images of the edge of a Ni grain at the critical steps of the reoxidation and reduction processes. Black arrows and arrowheads highlight key morphological changes occurring at the surface of the Ni grain, as discussed in the text.*

To verify that the second voltage increase coincides with the nucleation of Ni islands on the NiO scale, higher magnification images of the interface analysed in Figure 3a are reported in Figure 4. Lattice fringes can be periodically resolved, enabling an indexation of the different phases depending on the environmental conditions. At low $O_2$-to-$H_2$ ratios, fast Fourier transforms (FFT) reveal that the presence of Ni (111) reflections coincides with the first voltage increase observed after ~400 s (Figure 4a), in agreement with previous interpretations. When reaching higher $O_2$ partial pressures, new crystalline domains form on the surface of the Ni grains 550 s after the start of the experiment. Lattice fringes with the same lattice spacing (~4.7 $nm^{-1}$) but with a different orientation than the parent Ni grain can be resolved: these are attributed to NiO (200) planes (Figure 4c, g). In addition, faint reflections that correspond to NiO (111) planes can also be detected (4.1 $nm^{-1}$). At this point, the voltage starts to drop, which is consistent with an oxidation and hence deactivation of the Ni active sites in the anode. The $O_2$ flow was then stopped before fully oxidising the Ni grains. After an incubation time (from 1500 s to 2150 s), Ni (200) reflections (5.3 $nm^{-1}$) start to appear on the NiO scale (Figure 4d, h). This observation confirms the presence of Ni islands on the NiO scale in these intermediate $O_2$-to-$H_2$ ratio conditions. The nucleation of Ni on the surface of NiO enables the fuel to oxidize, which leads to an increase in voltage as long as $O_2$ reduces on the cathode side. The NiO scale becomes fully reduced after 2600 s as the $O_2$-to-$H_2$ ratio decreases: the Ni islands present on the surface eventually merge with the parent Ni grain (Figure 4e, i). As it was not fully oxidised, the Ni grain keeps its initial orientation after one partial oxidation and reduction cycle. In parallel, the voltage drops as the partial pressure of $O_2$ becomes insufficient to sustain the reduction of oxygen at the cathode. A movie of the full sequence of data shown in Figure 4 can be found in the supplementary information (Video S3).

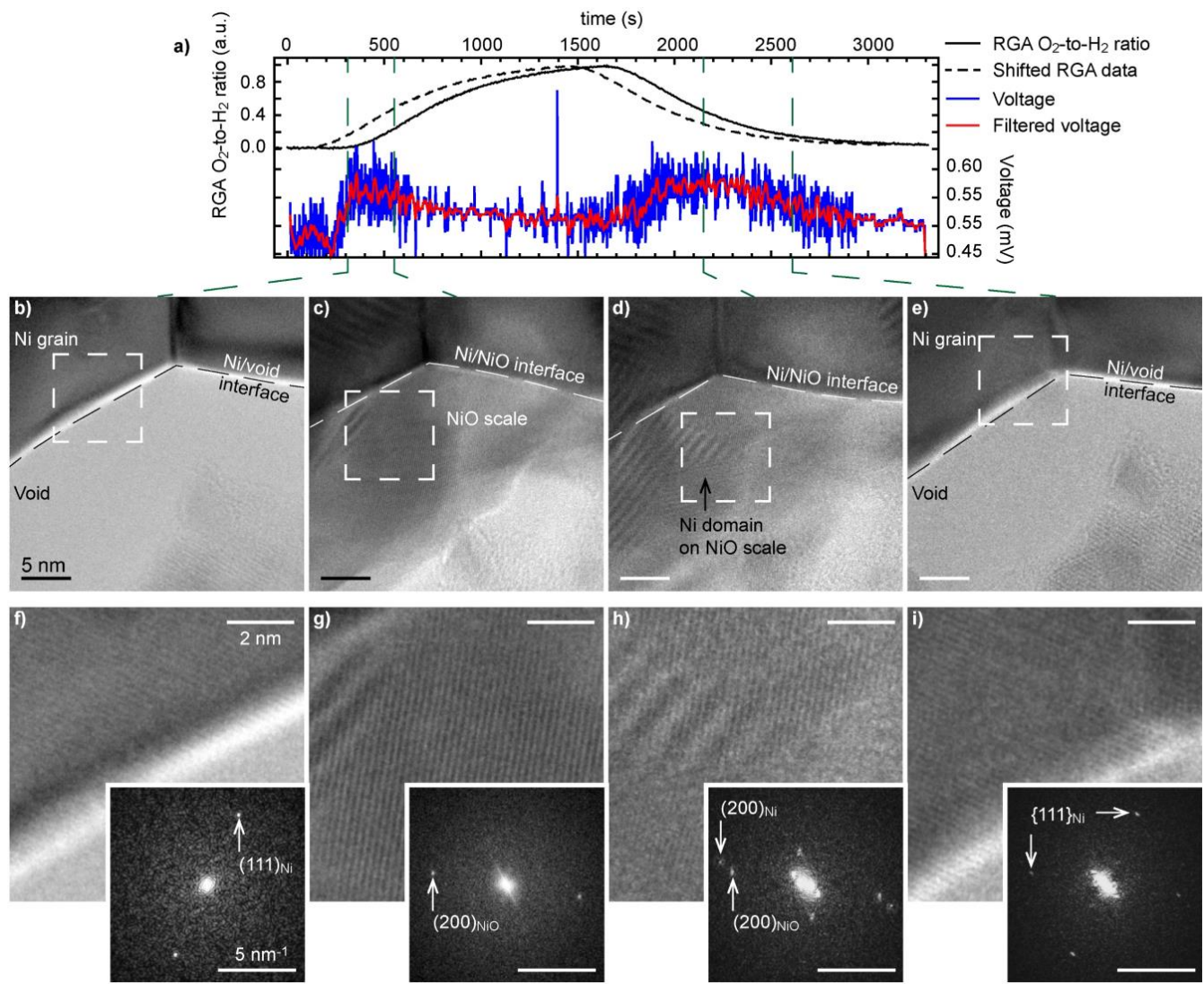

*Figure 4: a) RGA $O_2$-to-$H_2$ ratio (raw data, full line, and shifted forward in time by 180 seconds, dashed line), and voltage measured between the anode and cathode (raw and gaussian-filtered data). b-e) high-resolution TEM images of the edge of a Ni grain at the critical steps of the reoxidation and reduction processes, and f-i) Fourier-filtered magnified micrographs and corresponding FFTs taken from the dashed regions in b-e). The interface of the Ni grain, with a void or with the NiO scale, is marked by a dashed line in b-e).*

Overall, the minimum flow allowed by the mass flow controllers coupled with the maximum pressure achievable in the ETEM chamber limited us to transient experiments: a constant $O_2$-to-$H_2$ ratio maintaining the Ni reduced on the anode side and sufficient $O_2$ at the cathode to ensure a continuous cell operation could not be reached in the ETEM. In other words, the thin lamella oxidised quickly upon $O_2$ exposure. These oscillations in open circuit voltage were reproduced *ex situ* (Figure S4) and have also been reported in the literature with bulk systems.[46]

Finally, the cathode microstructure remained unchanged after redox cycling and cell operation in these conditions (Figure S2).

Figure 5 summarises our nanoscale observations of a SOFC obtained by *operando* ETEM in a single-chamber configuration. Starting with the Ni catalyst in its metallic electrochemically active state, the cell voltage remains low at small $O_2$-to-$H_2$ ratios as the $O_2$ reduction reaction at the cathode side is inhibited due to the absence of $O_2$ (Figure 5a). At intermediate $O_2$-to-$H_2$ ratios, the difference in catalytic activity between the anode and cathode ensures that $O_2$ reduces at the cathode and $H_2$ oxidizes at the anode, resulting in a small yet measurable voltage difference indicative of an operation of the cell (Figure 5b). As the $O_2$-to-$H_2$ ratio continues to increase, the Ni catalyst starts to oxidize on its surface, which inhibits the adsorption and dissociation of $H_2$, and hence stops the electrochemical reaction (Figure 5c). When $O_2$ is removed from the ETEM chamber and the $O_2$-to-$H_2$ ratio decreases, Ni islands start to nucleate on the NiO scale, leading to the formation of electrochemically active sites in the anode and to a voltage gain (Figure 5d). The NiO scale then completely reduces to metallic Ni as the $O_2$-to-$H_2$ ratio decreases further. Below a certain $O_2$-to-$H_2$ ratio threshold, the partial pressure of $O_2$ is insufficient to sustain the oxygen reduction reaction in the cathode and the $H_2O$ formation reaction stops (Figure 5e).

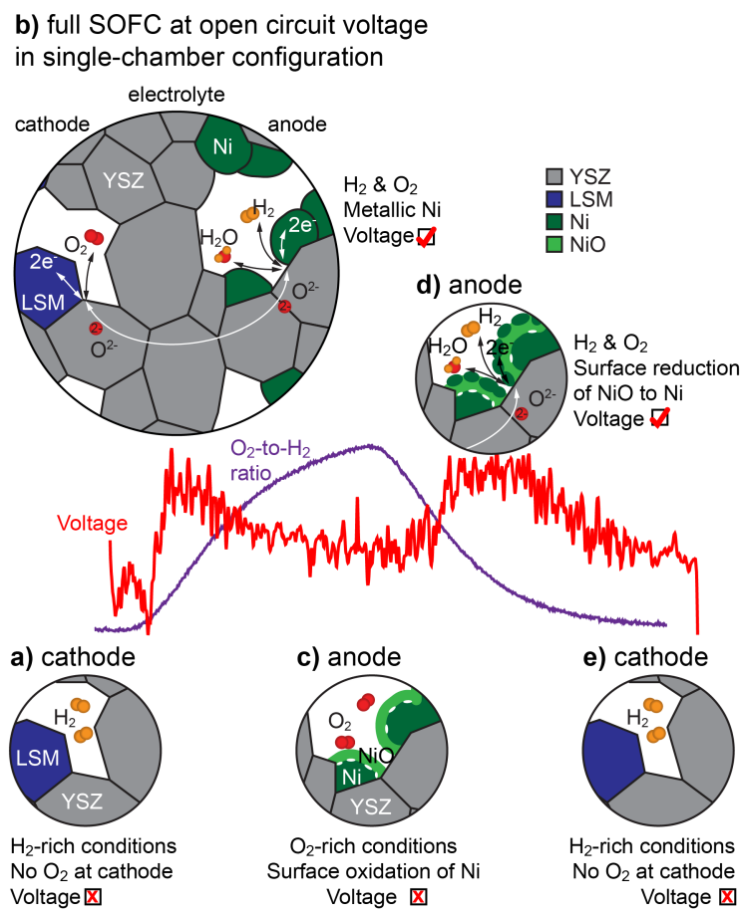

*Figure 5: Schematic summary of the operation of the SOFC in a single-chamber configuration as observed operando in the ETEM. a) At low $O_2$-to-$H_2$ ratios, the absence of $O_2$ prevents its reduction at the cathode. b) When introducing $O_2$, the cell starts to deliver a voltage synonym of its operation until c) the surface of the Ni grains oxidizes. d) When decreasing the $O_2$-to-$H_2$ ratio, the surface of the NiO scale starts to reduce into Ni islands, re-initiating the oxidation of the fuel at the anode, which results in a voltage increase. d) The process stops at low $O_2$-to-$H_2$ ratios (as in a).*

**Conclusions**

We demonstrated here that a SOFC can be analysed *operando* in a single-chamber configuration by ETEM. Both $H_2$ and $O_2$ were introduced in the microscope chamber, whilst keeping the cell at high operating temperature (600 °C) and observing its microstructure down to the atomic scale. By varying the $O_2$-to-$H_2$ ratio, direct correlations between cell voltage, gas atmosphere and microstructure of the Ni catalyst were established. At intermediate $O_2$-to-$H_2$ ratios and when the Ni catalyst is maintained in its metallic state, a small yet distinct gain in voltage between the two electrodes of the thin FIB-prepared lamella is measured. This open circuit voltage that builds up in these conditions results from the difference in fuel oxidation and oxidant gas reduction selectivity between the anode and cathode. Depending on $O_2$-to-$H_2$ ratio, the surface oxidation of Ni stops the fuel oxidation reaction, while the growth of Ni islands on the NiO scale restarts it. Looking ahead, such *operando* experiments in the ETEM should enable to investigate a wide range of degradation pathways affecting SOFCs, notably the poisoning of electrochemically active TPBs of both cathode and anode, or the impact of a coarsening of the Ni catalyst.

**Materials and methods**

FIB samples for *operando* characterization were prepared using a ZEISS Crossbeam 540 and contacted to a double-tilt 6 contacts DENSsolutions TEM holder and eventually to a voltmeter. The FIB-prepared samples were thinned once on the MEMS chip using a final voltage of 5 kV to reduce $Ga^+$-induced damages and possible $Ga^+$-rich surface short-circuits, which are particularly detrimental to biasing experiments. The thickness of the TEM lamella was ~200 nm. TEM experiments were performed in an image-Cs-corrected environmental FEI Titan microscope operated at 300 kV equipped with a CMOS camera (Gatan Oneview), a solid state EDX detector (Oxford 80), and an electron energy-loss spectrometer (Gatan Tridiem 965 ER). Analysis involved STEM imaging using an ADF detector, high-resolution TEM imaging, EDX and EELS. TEM movies were recorded using a home-made Gatan Digital Micrograph script, which blanks the electron beam between image acquisitions to avoid any contribution of secondary electrons to the measured voltage and beam-induced artefacts. TEM micrographs were acquired here every 20 seconds. A mass spectrometer (Pfeiffer Vacuum Model PrismaPlus™ QMG 220) located at the exit of the ETEM chamber was used to quantify the $O_2$-to-$H_2$ gas ratio by following the mass-to-charge ratios of 32 and 2 for $O_2$ and $H_2$, respectively. The time taken by the gas to reach the RGA was estimated by monitoring the delay between the introduction of the gas and its detection by the RGA. Data from the voltmeter was filtered: outliers induced by the periodic presence of the electron beam (every 20 seconds) were removed using a homemade *Mathematica* script, before filtering the resulting data using a gaussian filter spanning across 4 data points. EELS acquisitions were carried out in STEM mode using the spectrum imaging approach implemented in *Gatan Digital Micrograph.* The spectra were background-subtracted using a power law function, aligned with respect to the Ni-$L_3$ energy at ~855 eV, and spectra were normalized on the Ni-$L_2$ edge to highlight the evolution of the $L_3/L_2$ intensity ratio. The TEM lamellae were relatively thick (thickness/electron mean free path $\geq 1$) to maintain the structural integrity of the SOFC, however preventing a precise quantification of the Ni oxidation state.

*Ex situ* tests were performed using a 14-mm diameter button cell featuring the same materials as those tested by *operando* ETEM. The button cell was pressed between two gold meshes and placed inside a vertical furnace (Rohde, TE 10 Q SEV). The gas composition was adjusted by mixing individual gases. Each flow was accurately controlled by calibrated primary digital Mass Flow Controllers (MFCs, Bronkhorst, F-201CV, $\Delta\Phi = 0.005\Phi + 0.001$ max scale). Equal gas flows were sent to the anode and cathode, from the centre of the cell and spreading outwards. The cell was heated up to 600 °C at a rate of 25 °C/h under ambient atmosphere then purged with pure nitrogen. The reduction of the nickel anode was performed with 5 v/v% of $H_2$ in $N_2$ for 22 h, before ramping up and down the $O_2$ content of the gas. The cell voltage was measured between the two gold meshes and the temperature was measured with a K-type thermocouple placed as close as possible to the cell (about 1 mm). Both signals were recorded online with a data logger (Fluke, Hydra).

**Acknowledgements**

The authors acknowledge funding from the French microscopy network METSA (www.metsa.fr) for the ETEM access. The ETEM work was performed at the Consortium Lyon-St-Etienne de Microscopie (CLYM, www.clym.fr). M B., T.E. and M.D. acknowledge funding from the INSTANT project of the France-Singapore MERLION program 2019-2020. M.D. acknowledges the Facilities for Analysis, Characterization, Testing and Simulations (FACTS) at the Nanyang Technological University for access to the FIB equipment, as well as financial support from the Nanyang Technological University start-up grant (Grant M4081924). The authors also would like to thank Z. Wuillemin for inputs regarding the SOFC samples.

**Authors contribution**

Q.J. and M.D. designed the experiments. E.T. prepared the TEM lamellae from a bulk SOFC sample with a thinner YSZ electrode provided by D.M.. G.P. provided the TEM holder. M.B., T.E., M.D. and Q.J. performed the ETEM experiments and analysed the data. C.F. and S.D. performed the *ex situ* experiments. A.W.H. and J.V.H. supervised the research. Q.J. wrote the manuscript with the help of all co-authors.

*Supplementary information – Operando analysis of a solid oxide fuel cell by environmental transmission electron microscopy*

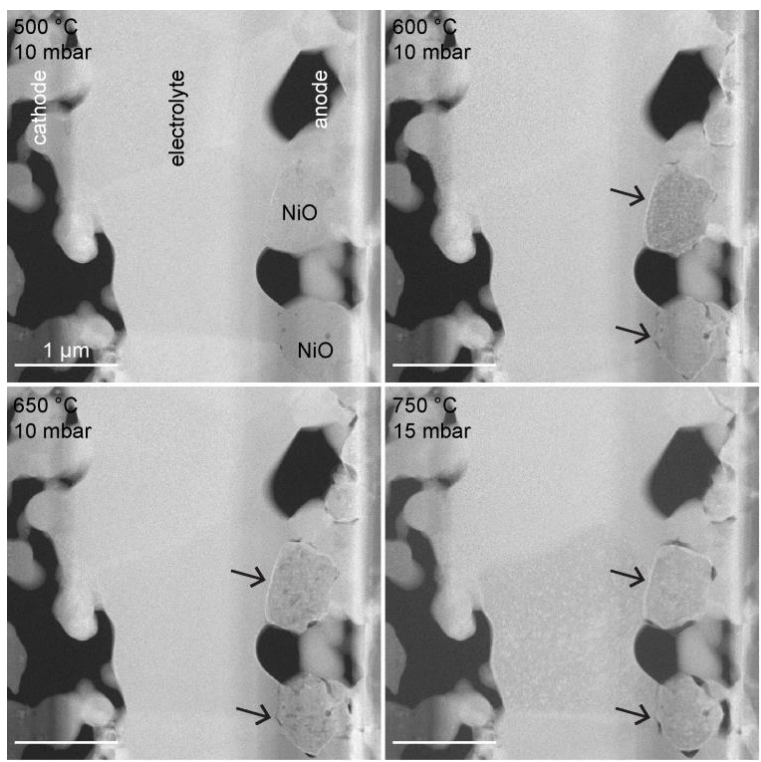

*Figure S1: Sequence of STEM ADF of the first reduction of the SOFC lamella to activate the Ni catalyst. Arrows highlight the changes in morphology of the NiO grains, which shrink upon removal of oxygen. The orientation of the sample changed upon heating, leading to a difference in contrast (especially in the electrolyte YSZ phase).*

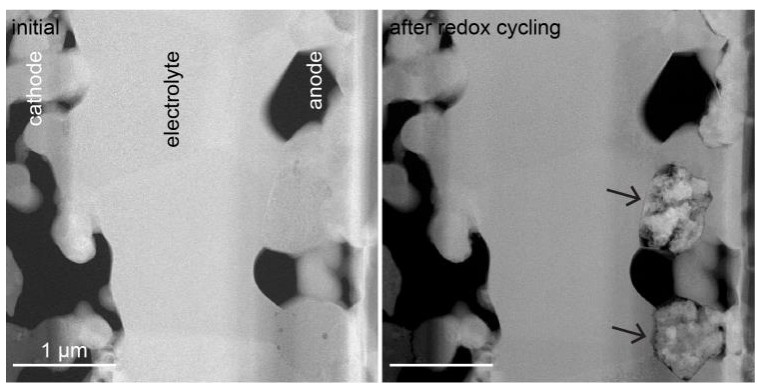

*Figure S2: STEM ADF images of the SOFC lamella before and after multiple redox cycling and cell operation. The arrows show two characteristic Ni grains. The cathode remained unchanged during these experiments.*

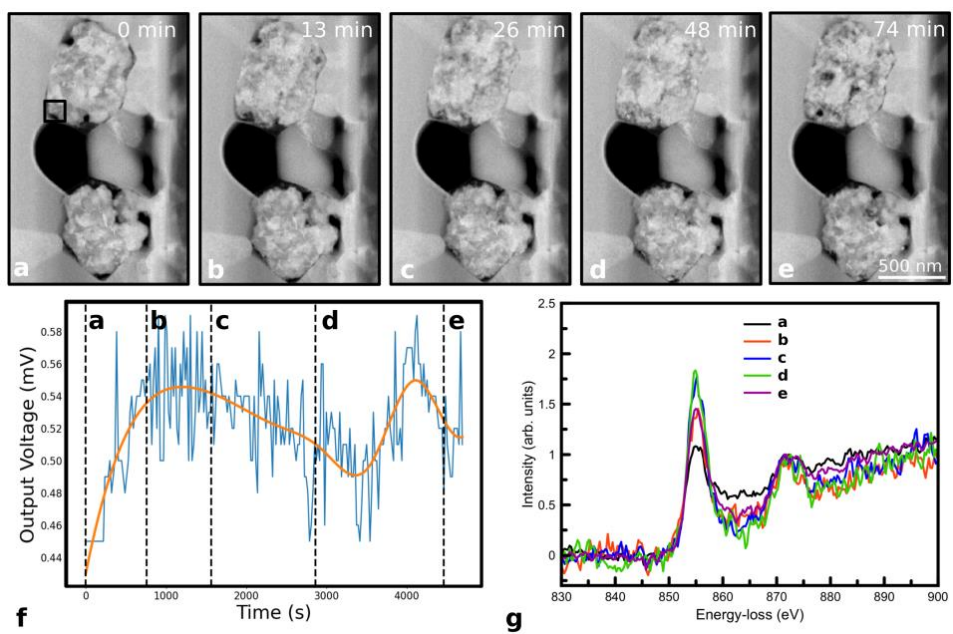

*Figure S3: a-e) STEM ADF micrographs of the Ni grains of the SOFC anode, first in 16 mbar of H$_2$/N$_2$, then exposed to an increasing amount of O$_2$ from 0 to 3000 s (a-d), before removing O$_2$ from the ETEM column 3000 s after the start of the experiment (e). A volume expansion occurs upon oxidation (a-d), while the Ni volume shrinks upon reduction (d-e). f) Corresponding evolution of the cell voltage with varying O$_2$-to-H$_2$ ratios (with and without filtering). The vertical dashed lines next to the letters a-e) correspond to the recording time of the corresponding micrographs a-e). g) EEL spectra of the Ni-L$_{2,3}$ edges recorded simultaneously with the STEM ADF micrographs shown in a-e) in the area marked by the black square in a). The spectra were background-subtracted, re-aligned in energy at the L$_3$ edge (855 eV) and normalized using the L$_2$ edge (872 eV). The increase in L$_3$-edge intensity with increasing O$_2$ concentration is visible, indicating an oxidation of Ni to NiO from a) to d).[47] The spectrum e) was measured after the removal of O$_2$ from the column and is representative of an intermediate NiO/Ni state, see main text for details.*

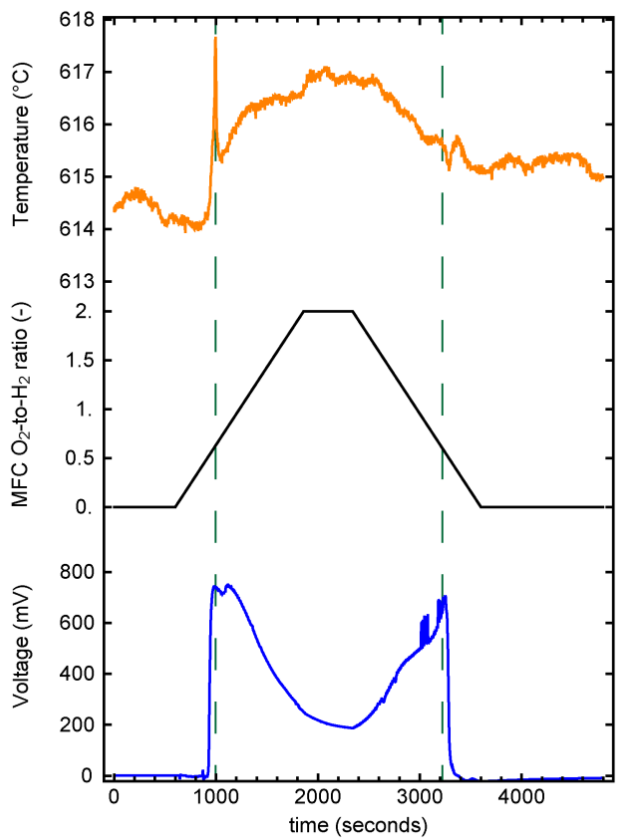

*Figure S4: Ex situ measurements of the evolution of the open circuit voltage and temperature together with the $O_2$-to-$H_2$ ratio (from calibrated MFCs) of a button cell featuring the same cathode-electrolyte-anode materials as those tested operando by ETEM. The cell operates around the stoichiometric $O_2$-to-$H_2$ ratio of 0.5.*

Movies of the full sequences of data shown in Figures 2,3 and 4 are provided as separate files (Videos S1, S2, S3, respectively)